\documentclass[twocolumn,aps,preprintnumbers,superscriptaddress]{revtex4}
\usepackage{graphicx}
\usepackage{dcolumn}
\usepackage{bm}
\usepackage{times}
\usepackage[colorlinks=true,linkcolor=blue,anchorcolor=blue,citecolor=blue,urlcolor=blue]{hyperref}
\usepackage{multirow}
\usepackage{amsmath}
\usepackage{amsfonts}
\usepackage{amssymb}

\begin{document}

\title{ Topological crystalline antiferromagnetic state in tetragonal FeS}
\author{Ningning Hao}
\thanks{Equal contributions}
\affiliation{Anhui Province Key Laboratory of Condensed Matter Physics at Extreme
Conditions, High Magnetic Field Laboratory, Chinese Academy of Sciences,
Hefei 230031, Anhui, China}
\affiliation{Collaborative Innovation Center of Advanced Microstructures, Nanjing
University, Jiangsu Province 210093, China}
\author{Fawei Zheng}
\thanks{Equal contributions}
\affiliation{Institute of Applied Physics and Computational Mathematics, Beijing 100088,
China}
\author{Ping Zhang}
\email{zhang\_ping@iapcm.ac.cn}
\affiliation{Institute of Applied Physics and Computational Mathematics, Beijing 100088,
China}
\affiliation{Beijing Computational Science Research Center, Beijing 100193, China}
\author{Shun-Qing Shen}
\email{sshen@hku.hk}
\affiliation{Department of Physics, The University of Hong Kong, Pokfulam Road, Hong
Kong, China}

\begin{abstract}
Integration between magnetism and topology is an exotic phenomenon in
condensed-matter physics. Here, we propose an exotic phase named topological
crystalline antiferromagnetic state, in which antiferromagnetism
intrinsically integrates with nontrivial topology, and we suggest such a
state can be realized in tetragonal FeS. A combination of first-principles
calculations and symmetry analyses shows that the topological crystalline
antiferromagnetic state arises from band reconstruction induced by pair
checker-board antiferromagnetic order together with band-gap opening induced
by intrinsic spin-orbit coupling in tetragonal FeS. The topological
crystalline antiferromagnetic state is protected by the product of
fractional translation symmetry, mirror symmetry, and time-reversal
symmetry, and present some unique features. In contrast to strong
topological insulators, the topological robustness is surface-dependent.
These findings indicate that non-trivial topological states could emerge in
pure antiferromagnetic materials, which sheds new light on potential
applications of topological properties in fast-developing antiferromagnetic
spintronics.
\end{abstract}

\maketitle


\section{Introduction}

Since topological insulators were discovered theoretically and
experimentally \cite{Hasan-RMP-2010,Qi-RMP-2011,Shen-book-12},
symmetry-protected topological phases have become a general principle to
explore exotic quantum states of matter among complex and rich compounds\cite%
{Schnyder-PRB-2008,Chen-Science-2012,Gu-PRB-2009}. Various combinations of
time-reversal symmetry, crystal space group symmetry and particle-hole
symmetry may result in a large number of exotic topological quantum states
of matter. Among these, one example is the topological crystalline
insulator, in which the nontrivial topological properties are protected by
point group symmetry such as rotation, reflection, mirror symmetry, etc\cite%
{Fu-PRL-2011,Hsieh-NC-2012}, and the surface boundary that preserves the
underlying point group symmetry hosts the gapless surface states. To
fabricate a device with topological crystalline insulators, manipulation of
the spin degrees of freedom of surface electrons is essential. One possible
method is to include magnetism. For example, by doping magnetic atoms to
induce ferromagnetism, the quantum anomalous Hall effect can be realized in
a magnetic topological insulator\cite{Yu-Science-2010, Chang-Science-2013}.
However, doping magnetic atoms in topological state is very tough and
usually requires exquisite experimental designs. If the topological states
of matter possess intrinsic magnetism, there would be more room to
manipulate quantum spin of surface electrons.

In this work, we extend the concept of topological crystalline insulators
from nonmagnetic materials to antiferromagnetic materials. We demonstrate
that tetragonal FeS could be in a topological crystalline antiferromagnetic
state in the spirit of symmetry-protected topological phases. The nontrivial
topological crystalline antiferromagnetic state is protected by a
combination of fractional translation, mirror reflection, and time-reversal
symmetry. The fractional translation symmetry is induced by the pair
checker-board antiferromagnetic order instead of specific lattice structure%
\cite{Wang-nature-2016,Chang-NP-2017}. As a consequence, it is found that
the topological crystalline antiferromagnetic state has robust gapless
surface states on the crystal (010) surfaces, while on other surfaces, such
as (100) and (001) surfaces, there are no robust gapless surface states due
to the glide-plane mirror symmetry breaking. The existence of these surface
states is dictated by a mirror Chern number\cite{Teo-PRB-2007,Hsieh-NC-2012}%
. Therefore, tetragonal FeS is an ideal candidate with integration of
antiferromagnetism and topology, and provides a playground to study the
intrinsic magnetic effect on the surface states of topological crystalline
insulators. Furthermore, in comparison with nonmagnetic and ferromagnetic
materials, the antiferromagnetic topological materials have many advantages,
and attract more attentions\cite%
{Barker-PRL-2016,Zhang-SR-2016,Jungwirth-NN-2016}. Also, the tetragonal FeS,
by itself, is a kind of unconventional superconductor at low temperatures%
\cite{Lai-JACS-2015}. Thus the material provides an intrinsic platform to
study the interplay between topology, magnetism and superconductivity.

\section{The antiferromagnetic order}

Tetragonal FeS has a simple anti-PbO structure as shown in Fig. \ref%
{fig_lattice} (a). It has attracted great attention since the firstly
reported superconductivity with transition temperature 4.5K\cite%
{Lai-JACS-2015}. One of the important aspects of tetragonal FeS is to
identify the possible magnetic-ordered state in the vicinity of
superconductivity. To date, some magnetic states have been proposed for
tetragonal FeS experimentally, such as nonmagnetic metallic state, a
commensurate antiferromagnetic order with wave vector $k_{m}=(0.25,0.25,0)$,
low-moment ($10^{-2}-10^{-3} \mu_{B}$) ferromagnetic state coexisting with
superconductivity, and high-moment (about $1 \mu_{B}$) ferromagnetic state
coexisting with superconductivity\cite%
{Borg-PRB-2016,Kuhn-Ph-2016,Holenstein-PRB-2016,Kirschner-PRB-2016}. These
sample-dependent inconsistencies may need further efforts to be devoted into
the high-quality single crystal synthesis and relevant thin film growth.

In comparison with its widely studied isostructures like FeSe and FeTe,
first-principles calculations provide an effective method to identify the
magnetic ground state of tetragonal FeS by means of determination of the
lowest energy among all possible magnetic-ordered states. Possible
magnetic-ordered states of tetragonal FeS include the paramagnetic order,
collinear (single stripe), checker-board (N\'{e}el) and pair checker-board
(stagger dimer) antiferromagnetic orders.

\begin{figure}[tp]
\begin{centering}
\includegraphics[width=1.0\linewidth]{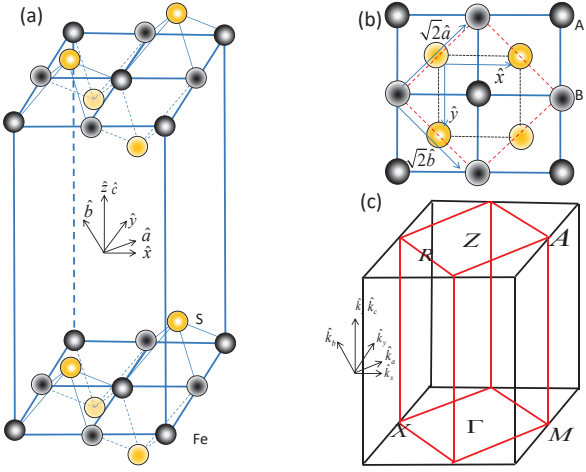}
\par\end{centering}
\caption{ (a)The 3D view of crystal structure of tetragonal FeS. Different
filling balls denote different sublattices. (b) The top-down view of
tetragonal FeS with the patterns of the paramagnetic order. The black-dotted
and red-dashed lines label the one-Fe and two-Fe unit cells. (c) Bulk
Brillouin zones (BZ) of (a). Here, we take the one-Fe-unit-cell constant as
length scale. $(\hat{x},\hat{y},\hat{z})$, $(\hat{a},\hat{b},\hat{c})$, $(%
\hat{k}_{x},\hat{k}_{y},\hat{k}_{z})$, and $(\hat{k}_{a},\hat{k}_{b},\hat{k}%
_{c}) $ denote the unit direction vectors in lattice and momentum space,
respectively. }
\label{fig_lattice}
\end{figure}

The density functional theory (DFT) and hybrid functional calculations in
the present work were performed by using \textit{Vienna ab-initio Software
Package} (VASP) \cite{vasp}. The atom core electrons were described by the
projector augmented wave (PAW) method \cite{paw1,paw2}.
Perdew-Burke-Ernzerhof (PBE) functional \cite{pbe1} was used to treat the
electronic exchange correlation. The energy cutoff for the plane-wave basis
was set to be 400 eV. The first Brillouin zone was sampled in the k-space
with Monkhorst-Pack scheme and the grid sizes are 19$\times $19$\times $13,
19$\times $19$\times $13, 13$\times $13$\times $13, and 13$\times $7$\times $%
13 for PM, C-AFM, COL-AFM, and PCB-AFM phases, respectively. We have checked
that the total energy is converged for the cutoff energy and the k-point
sampling. The atomic structure was relaxed until the force on each atom is
smaller than 0.01eV/\AA . Considering that there exists weak interlayer
coupling in this composite system, we have added the van der Waals
correction to the DFT calculations \cite{vdw}. In order to take into account
the correlation from a moderate Hubbard $U$ interaction, generalized
gradient approximation (GGA)+$U$ calculation method is used.

\begin{figure}[tp]
\begin{centering}
\includegraphics[width=1.0\linewidth]{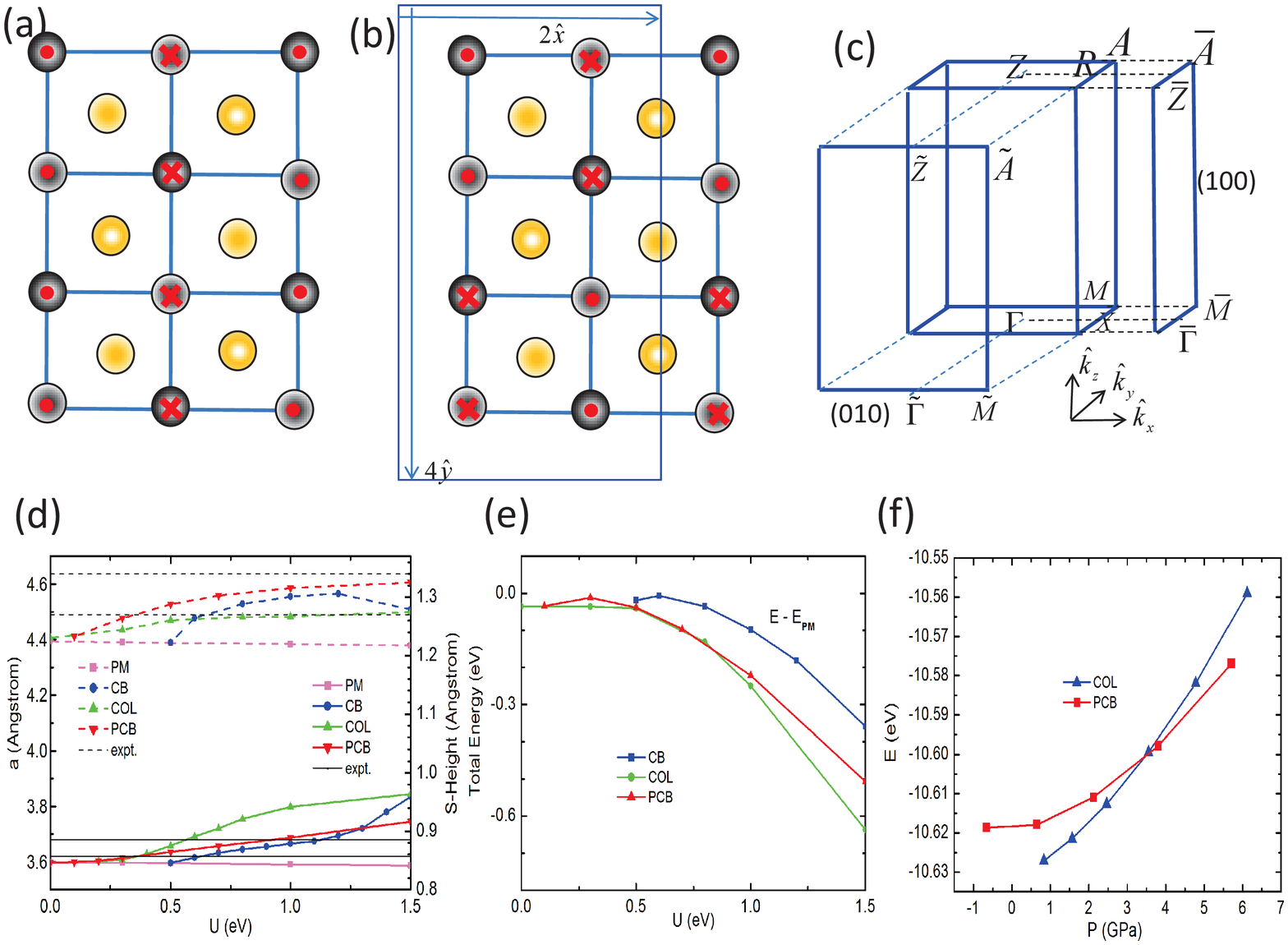}
\par\end{centering}
\caption{(a) and (b) The top-down view of tetragonal FeS with the patterns
of the COL-AFM order in (a) and the PCB-AFM order in (b). The red spots and
red arrows denote the magnetization outward and inward the iron plane. In
(b) the solid-blue lines label the unit cell of the PCB-AFM phase. (c)
Middle: bulk Brillouin zone of the PCB-AFM state; Left: (010)-surface
Brillouin zone; Right: (100)-surface Brillouin zone. Here, we also take the
one-Fe-unit-cell constant in the paramagnetic state as length scale. $(\hat{x%
},\hat{y},\hat{z})$, $(\hat{a},\hat{b},\hat{c})$, $(\hat{k}_{x},\hat{k}_{y},%
\hat{k}_{z})$, and $(\hat{k}_{a},\hat{k}_{b},\hat{k}_{c})$ denote the unit
direction vectors in lattice and momentum space, respectively. (d) The
in-plane lattice constants and the S atom heights to iron plane as functions
of Hubbard U modulation in different magnetic phases at ambient pressure.
Here, PM, C-AFM label paramagnetic and checker-board antiferromagnetic
states, respectively. (e) The total energies per Fe$_{2}$S$_{2}$ as
functions of Hubbard U modulation in different magnetic phases at ambient
pressure. In both (d) and (e), the results are from GGA+U method.}
\label{fig_LDA1}
\end{figure}

The magnetization patterns of the collinear antiferromagnetic (COL-AFM) and
pair checker-board antiferromagnetic (PCB-AFM) states are shown in Fig. \ref%
{fig_LDA1}(a), (b). At ambient pressure, the lattice constant of the iron
plane and the S heights to the iron plane as functions of the Hubbard U are
shown in Fig.\ref{fig_LDA1}(c). To determine the suitable Hubbard U in
tetragonal FeS, we use a simple method. We start with the free lattice
constants. When Hubbard U is turned on and increases from zero, the lattice
sites relax freely and achieve the equilibrium positions finally. We use the
obtained lattice constants to compare with the experimentally measured
values. The suitable Hubbard U is read out when the two sets of lattice
constants match with each other. It can be seen that the Hubbard U with 1eV
is reasonable to match the experimental lattice parameters. Under such a
Hubbard U interaction correction, the COL-AFM and PCB-AFM states compete
with each other as shown in Fig.\ref{fig_LDA1}(d). The first-principles
calculations show that the COL-AFM state has a lower energy of 14meV/Fe than
the PCB-AFM state under the Hubbard U interaction correction of 1eV at
ambient pressure. Likewise, under ambient conditions, the recent neutron scattering
experiment observed the peaks of spin excitation at wave vector ($\pi ,0$),
which was the same wave vector of COL-AFM order\cite{Man-njp-2017}. The
predictions from the first-principles calculations follow the experimental
observations under ambient conditions. However,
first-principles calculations predict that the PCB-AFM state becomes the
lowest energy state with increasing pressure over the threshold value. For instance, the
energy in the PCB-AFM state is about 9mev/Fe lower than the energy in the
COL-AFM state at pressure of 4GPa. Note that the switching between different
orders tuned by pressure also occurs in bulk FeSe, in which the pressure
over 2GPa can change the state of FeSe from the nematic order to long-range
stable antiferromagentic order\cite{Sun-NC-2016,Sun-PRL-2017}. If the spin
excitation of the PCB-AFM order was identified by neutron scattering
measurement in tetragonal FeS under high pressure conditions, it would be
benefit not only to understand the superconductivity but to study the
topological states in tetragonal FeS. Interestingly, the PCB-AFM state is
predicted to be the ground state in many other iron chalcogenides such as
FeSe, monolayer FeSe and pressured FeSe\cite%
{Cao-PRB-2015,Liu-PRB-2016,Glasbrenner-NP-2015}.

The low-energy states with various magnetic orders in iron-based materials
can be captured by a minimal Heisenberg $J_{1}-J_{2}-J_{3}-K$ spin model
\cite{Hu-PRB-2012},

\begin{equation}
H=\sum_{nn}[J_{1}\hat{S}_{i}\cdot \hat{S}_{j}-K(\hat{S}_{i}\cdot \hat{S}%
_{j})^{2}]+\sum_{2nn}J_{2}\hat{S}_{i}\cdot \hat{S}_{j}+\sum_{3nn}J_{3}\hat{S}%
_{i}\cdot \hat{S}_{j}.  \label{Heisenberg}
\end{equation}%
Here, $nn$, $2nn$ and $3nn$ denote the nearest, second nearest, and third
nearest neighbor, respectively. The mean-field phase diagram for the
Hamiltonian in Eq. (\ref{Heisenberg}) was presented in Fig. 2 in reference%
\cite{Glasbrenner-NP-2015}. For tetragonal FeS, the first-principles
calculations give the model parameters shown in Table I. It is
straightforward to check that the data in the first group give the magnetic
ground state with the COL-AFM order while the data in the second group give
the magnetic ground state with the PCB-AFM order, according to the phase
diagram in Fig. 2c in reference\cite{Glasbrenner-NP-2015}. Thus, we propose
that the state with the PCB-AFM order is the magnetic ground state in
tetragonal FeS at 4GPa. As we mentioned in the introduction, the PCB-AFM
order can induce an uniqe topological crystalline antiferromagnetic state.
In the remaining part of the paper, we focus on the discussons how the
PCB-AFM order drives the topological crystalline antiferromagnetic state and
about the properties of the topological state. The properties of other
antiferromagnetci states are discussed in Supplemental Materials (SMs).
\begin{table}[tbp]
\caption{The calculated parameters of tetragonal FeS for $J_{1}$-$J_{2}$-$%
J_{3}$-$K$ model. }%
\begin{tabular}{|l|l|l|l|l|}
\hline\hline
$J_{1}$ & $J_{2}$ & $J_{3}$ & $K$ & Pressure \\ \hline
127.5 & 84.6 & -0.4 & 38.9 & at ambient pressure \\ \hline
126.1 & 71.6 & 14.0 & -8.2 & at 4GPa \\ \hline
\end{tabular}%
\end{table}

\section{Band reconstruction from PM state to PCB-AFM state}

\begin{figure}[tp]
\begin{centering}
\includegraphics[width=1.0\linewidth]{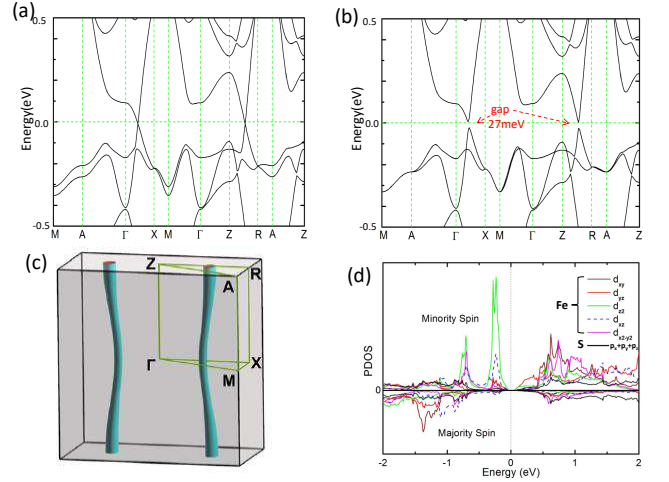}
\par\end{centering}
\caption{ (a) The band structures for the PCB-AFM state with the {[}100{]}%
-direction magnetization in the presence of spin-orbit coupling. (b) The
band structures for the PCB-AFM state with the {[}001{]}-direction
magnetization in the presence of spin-orbital coupling. (c) 3D Fermi surface
corresponds to (a) when the Fermi level is slightly shifted from zero. (d)
The partial density of states for $d$ orbitals of Fe and $p$ orbitals of S
in the PCB-AFM state with the {[}001{]}-direction magnetization in the
absence of spin-orbit coupling.}
\label{fig_band_LDA}
\end{figure}

Stable magnetic order in materials usually breaks some spatial symmetries
such that the underlying electronic structures are strongly reconstructed
with new features. It is widely known that the reconstructed band structures
host Dirac cone structures in BaFe$_{2}$As$_{2}$ with the collinear AFM order%
\cite{Richard-PRL-2010}. The reconstructed bulk band structures for
tetragonal FeS with the PCB-AFM order in the presence of spin-orbit coupling
are shown in Figs. \ref{fig_band_LDA}(a) and \ref{fig_band_LDA}(b), which
correspond to magnetization along the [100] and [001] directions,
respectively. The remarkable feature in Fig. \ref{fig_band_LDA}(a) is the
emergence of Dirac points in the $k_{z}=0$ and $\pi /c$ planes. Slightly
finite electron or hole doping can shift the Fermi level away from the Dirac
points, and the Fermi surfaces form two thin tubes in three dimensional
momentum space as shown in Fig. \ref{fig_band_LDA}(c). It means the PCB-AFM
state with the {[}100{]}-direction magnetization belongs to node-line
semimetal in the undoped case even in the presence of spin-orbit coupling.
Further calculations show that other PCB-AFM states with in-$xy$-plane
magnetization, such as {[}010{]} and {[}110{]} directions, have similar
results with the state with {[}100{]}-direction magnetization. However, no
Dirac points survive and a fully-gapped state is obtained in the PCB-AFM
state with the {[}001{]}-direction magnetization in the presence of
spin-orbit coupling.

To understand the band reconstruction of tetragonal FeS with the PCB-AFM
order, we introduce a tight-binding model associated with mean-field
approximation involving five $d$ orbitals of Fe by ignoring $p$ orbitals due
to their negligible weight around the fermi level shown in Fig. \ref%
{fig_band_LDA}(d),
\begin{equation}
H^{(l)}=H_{0}+H_{pcb}^{(l)}+H_{so},  \label{Htot}
\end{equation}%
where
\begin{equation}
H_{0}=\sum_{k\in BZ_{pm}}\sum_{\alpha ,\beta ,\sigma }\Psi _{\alpha ,\sigma
}^{\dag }(\mathbf{k})H_{\alpha \beta }(k)\Psi _{\beta ,\sigma }(\mathbf{k}),
\label{H_tb}
\end{equation}%
\begin{equation}
H_{pcb}^{(l)}=\sum_{k_{n}\in BZ_{pcb}}\sum_{n=0,\alpha }^{3}\Psi _{\alpha
}^{\dag }(\mathbf{k}_{n})\Delta e^{-i\theta (\alpha )}s_{l}\Psi _{\alpha }(%
\mathbf{k}_{n+1})+H.c.],  \label{H_PCB}
\end{equation}%
and

\begin{align}
H_{so}& =\sum_{k\in BZ_{pm}}\sum_{\alpha }[\Psi _{\alpha }^{\dag }(\mathbf{k}%
)\lambda _{so,z}L^{z}s_{z}\Psi _{\alpha }(\mathbf{k})  \notag \\
& +\Psi _{\alpha }^{\dag }(\mathbf{k})\lambda _{so,\parallel
}(L^{x}s_{x}+L^{y}s_{y})\Psi _{\bar{\alpha}}(\mathbf{k})].  \label{H_SOC}
\end{align}%
Here, $H_{0}$ is the tight-binding Hamiltonian describing the electronic
structure in the absence of magnetic orders. $\mathbf{k=}(k_{x},k_{y},k_{z})$
is defined in a one-Fe unit cell with $(k_{x},k_{y})\in \lbrack 0,2\pi ]$, $%
k_{z}\in \lbrack 0,2\pi /c]$. We have transformed $H_{0}$ from the
two-Fe-unit-cell representation to the one-Fe-unit-cell representation
according to the parity of the glide-plane symmetric operator $\frac{1}{2}%
\hat{t}(\sqrt{2}\hat{a},\sqrt{2}\hat{b},0)\hat{M}_{z}$ \cite%
{Hao-PRX-2014,Hao-PRB-2015}, with $\hat{t}$ and $\hat{M}_{z}$ the
translation operation and mirror reflection about $xy$-plane, respectively. $%
\alpha $ and $\beta $ take $o$ or $e$ to label the parity of the glide-plane
symmetry, and $\sigma $ labels the spin degrees of freedom. $\Psi _{o,\sigma
}^{T}(\mathbf{k})\mathtt{=}[d_{xy,\sigma }(\mathbf{k+Q})$, $%
d_{x^{2}-y^{2},\sigma }(\mathbf{k+Q})$, $d_{xz,\sigma }(\mathbf{k})$, $%
d_{yz,\sigma }(\mathbf{k})$, $d_{z^{2},\sigma }(\mathbf{k+Q})]$ and $\Psi
_{e,\sigma }(\mathbf{k})\mathtt{=}\Psi _{o,\sigma }(\mathbf{k+Q})$ with $%
\mathbf{Q=(}\pi ,\pi ,0\mathbf{)}$ denoting the folding wave vector from the
one-Fe BZ to the two-Fe BZ as shown in Fig. \ref{fig_lattice}(c). The exact
expressions for $H_{\alpha \beta }(k)$ are presented in SMs. $H_{pcb}^{(l)}$
describes the PCB-AFM order under the mean-field approximation. In the
PCB-AFM state, one magnetic unit cell includes eight Fe atoms as shown in
Fig. \ref{fig_LDA1}(b), and the corresponding folded wave vector is $\mathbf{%
Q}_{1}\mathtt{=}(\pi ,\pi /2,0)$. $\mathbf{k}_{n}\mathtt{=}\mathbf{k}\mathtt{%
+}n\mathbf{Q}_{1}$. $l\mathtt{=}0,x,y,z$ label the $2\times 2$ unit matrix
and Pauli matrices, respectively. $\Delta e^{-i\theta (\alpha )}\mathtt{\ }$%
labels the PCB-AFM order parameter with $\Delta \mathtt{=}$diag{[}$%
m_{xy},m_{x^{2}-y^{2}},m_{xz},m_{yz},m_{z^{2}}${]} and $\theta (\alpha
)=[n/2-(-1)^{\alpha }]\pi $. $H_{so}$ is the spin-orbit coupling term. $\bar{%
\alpha}$ denotes the inverse parity of $\alpha $.

\begin{figure}[tp]
\begin{centering}
\includegraphics[width=1.0\linewidth]{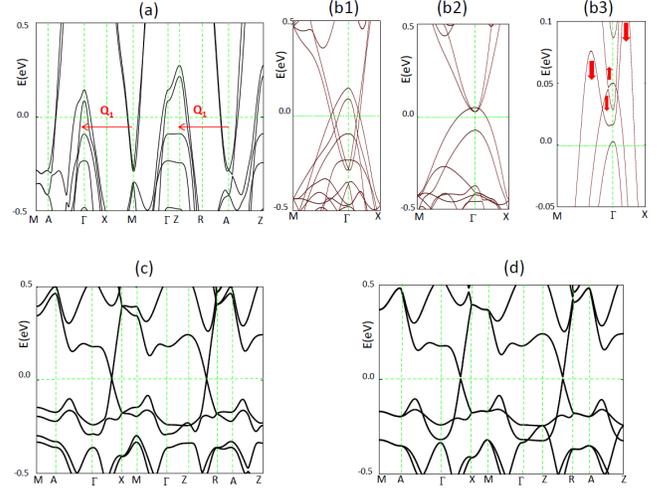}
\par\end{centering}
\caption{(a) Band structures in the paramagnetic state in the presence of
spin-orbit coupling. (b1)-(b3), (c)-(d) Bands evolution from the
paramagnetic state to the PCB-AFM state along the $M-\Gamma -X$ lines from
the mean-field Hamiltonian in Eq. (\protect\ref{Htot}) for the PCB-AFM state
with the {[}001{]}-direction magnetization in the presence of spin-orbit
coupling. (b1) Bands of the paramagnetic state in the folded BZ of the
PCB-AFM state. (b2) Turn on all $\Delta \protect\varepsilon $ and turn off
all $m$. (b3) Turn on all $\Delta \protect\varepsilon $ and turn off only $%
m_{xy}$. (c) Turn on all $\Delta \protect\varepsilon $ and all $m_{xy}$ to
obtain bands in the PCB-AFM state. We set $\protect\lambda _{so,\parallel }%
\mathtt{=}0.03$ eV, $\protect\lambda _{so,z}\mathtt{=}0.00$ eV in (c) and $%
\protect\lambda _{so,\parallel }\mathtt{=}0.00$ eV, $\protect\lambda _{so,z}%
\mathtt{=}0.03$ eV in (d). The modulations of the orbital energy (in unit of
eV) about the paramagnetic state and the PCB-AFM order parameters (in unit
of $\protect\mu _{B}$) take the magnitudes as follows, $\Delta \protect%
\varepsilon _{xy}\mathtt{=}\mathtt{-}0.375$, $\Delta \protect\varepsilon %
_{x^{2}-y^{2}}\mathtt{=}\mathtt{-}0.075$, $\Delta \protect\varepsilon _{xz}\mathtt{=}%
\mathtt{-}0.425$, $\Delta \protect\varepsilon _{yz}\mathtt{=}\mathtt{-}0.125$, $%
\Delta \protect\varepsilon _{z^{2}}\mathtt{=}0.025$; $m_{xy}\mathtt{=}1.04$%
, $m_{x^{2}-y^{2}}\mathtt{=}0.33$, $m_{xz}\mathtt{=}0.25$, $m_{yz}\mathtt{=}%
0.64$, $m_{z^{2}}\mathtt{=}0.30$. The sum of five $m$ is about 2.5 $\protect%
\mu _{B}$ as the result obtained from the first-principles calculations.}
\label{fig_band_TB}
\end{figure}

\begin{figure}[tp]
\begin{centering}
\includegraphics[width=1.0\linewidth]{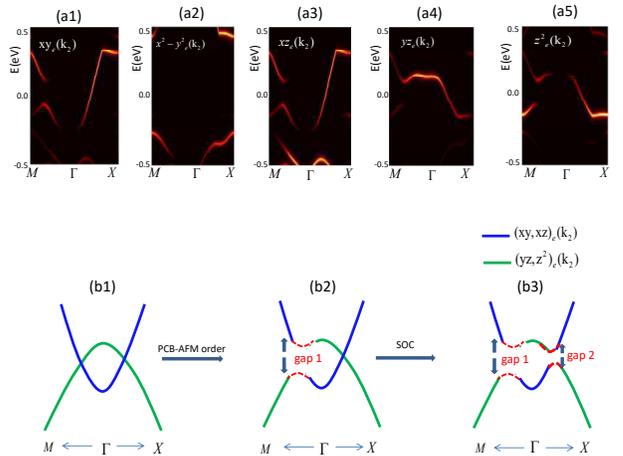}
\par\end{centering}
\caption{(a1)-(a5) The orbital-resolved spectral function $A_{e}(\mathbf{k}%
_{2},\protect\omega )$ for five $d$ orbitals from the mean-field Hamiltonian
in Eq. (\protect\ref{Htot}) in the PCB-AFM state with the {[}001{]}%
-direction magnetization. (b1)-(b3) Under the two effective bands picture,
the schematic diagrams show the topological phase transition induced by the
PCB-AFM order and spin-orbit coupling in tetragonal FeS.}
\label{fig_band_TB2}
\end{figure}
Fig.\ref{fig_band_TB}(a), (b1)-(b3), (c) demonstrate the band reconstruction
from the paramagnetic state to the PCB-AFM state. Starting from the band
structure in Fig.\ref{fig_band_TB}(a), the wave vector $\mathbf{Q}_{1}$
connects electron-type band near the $M$ ($A$) point with hole-type band
near the $\Gamma $ ($Z$) point, and folds one to another in the folded BZ of
the PCB-AFM state as shown in Fig.\ref{fig_band_TB}(b1). Define the orbital
energy modulations $\Delta \varepsilon _{l}$,
\begin{equation}
\Delta \varepsilon _{j}=e_{j}^{(PCB)}-e_{j}^{(PM)},
\label{det_orbital_energy}
\end{equation}%
where $e_{j}^{(PCB)}$ and $e_{j}^{(PM)}$ denote the energy of orbital
indexed by $j$ in PCB-AFM and PM states, respectively. $\Delta \varepsilon $
is induced by nematicity of the PCB-AFM state, \textit{i.e.}, the $C_{4}$
rotation symmetry connecting the [100] and [010] directions is broken in the
PCB-AFM state. The specific values of all $\varepsilon _{j}^{(PCB)}$ and $%
\varepsilon _{j}^{(PM)}$ are listed in SMs. When $\Delta \varepsilon _{j}$
is considered, only three bands are revealed to play a key role around the
Fermi level as in Fig.\ref{fig_band_TB}(b2). After all the PCB-AFM order
parameters $m_{j}$ except $m_{xy}$ are turned on, the three bands are
strongly modulated and a small band gap is opened between an electron band
and a hole band along the $\Gamma -M$ direction, but no band gap opens along
the $\Gamma -X$ direction in Fig.\ref{fig_band_TB}(b3). When $m_{xy}$
increases from zero, the three bands move along the directions marked by the
arrows in Fig.\ref{fig_band_TB}(b3). Finally, the nodel-line semimetal phase
emerges as shown in Fig.\ref{fig_band_TB}(c).

Note that the band structure for the PCB-AFM state with the {[}001{]}%
-direction magnetization in Fig. \ref{fig_band_TB}(c) has nonzero spin-orbit
coupling $(L^{x}s_{x}+L^{y}s_{y})$ and the band structure in Fig. \ref%
{fig_band_TB}(d) has nonzero spin-orbit coupling $L^{z}s_{z}$ with $%
L^{\alpha }$ and $s_{\alpha }$ ($\alpha =x,y,z$) labelling the relevant
matrices under orbital and spin basis, respectively (See SMs for
details). It clearly indicates that the fully-gapped phase is induced by the
term of $L^{z}s_{z}$. Indeed, this is the main reason for the
magnetization-direction-dependent fully-gapped phase. The space group is
non-symmorphic \textit{P}4/\textit{nmm }in the paramagnetic state for the
crystal structure of tetragonal FeS shown in Fig. \ref{fig_lattice}(a). The
five $d$ orbitals can be divided into two orthogonal subgroups \{$%
d_{xz},d_{yz}$\} and \{$d_{xy},d_{x^{2}-y^{2}},d_{z^{2}}$\} according to the
eigenvalues of the non-symmorphic operator$\frac{1}{2}\hat{t}(\sqrt{2}\hat{a}%
,\sqrt{2}\hat{b},0)\hat{M}_{z}$. For the three $t_{2g}$ orbitals \{$%
d_{xz},d_{yz},d_{xy}$\}, only the term of $L^{z}s_{z}$ induces coupling
between two orbitals in the same subgroup \{$d_{xz},d_{yz}$\}. Such a
coupling breaks the Dirac points and results in a fully-gapped state in the
PCB-AFM state with the {[}001{]}-direction magnetization (non-zero $\langle
s_{z}\rangle $). The model Hamiltonian in Eq. \ref{Htot} gives an explicit
description about the band reconstruction of tetragonal FeS with the PCB-AFM
order.

According to the aforementioned analyses, the magnetic structures can be
well described by the effective two-band model. Thus, we can construct a
simple model to summarize the band reconstruction induced by the PCB-AFM
order and the spin-orbit coupling. To this end, we first plot the
orbital-resolved spectral function $A_{e}(\mathbf{k}_{2},\omega )$ of the
PCB-AFM state with the {[}001{]}-direction magnetization in the absence of
the spin-orbit coupling in Fig. \ref{fig_band_TB2}(a1)-(a5). It clearly
shows that the four $d$ orbitals of iron can be divided into two groups to
form two sets of the effective bands shown in Fig. \ref{fig_band_TB2}(b1).
Note that the band inversion condition is natural due to the folding induced
by the PCB-AFM wave vector $\mathbf{Q}_{1}$ through comparing Fig. \ref%
{fig_band_TB2}(b1) with Fig. \ref{fig_band_TB}(a), (b1). After the PCB-AFM
order is turned on, the two bands couple with each other and open a gap
along the $\Gamma -M$ direction but not along the $\Gamma -X$ direction as
shown in Fig. \ref{fig_band_TB2}(b2). Finally, the Dirac node is fully
gapped along the $\Gamma -X$ direction, because the spin-orbit coupling only
concurs with the magnetization along the [001] direction as shown in Fig. %
\ref{fig_band_TB2}(b3).

\section{Topological robust surface states and topological invariants}

\begin{figure}[tp]
\begin{centering}
\includegraphics[width=1.0\linewidth]{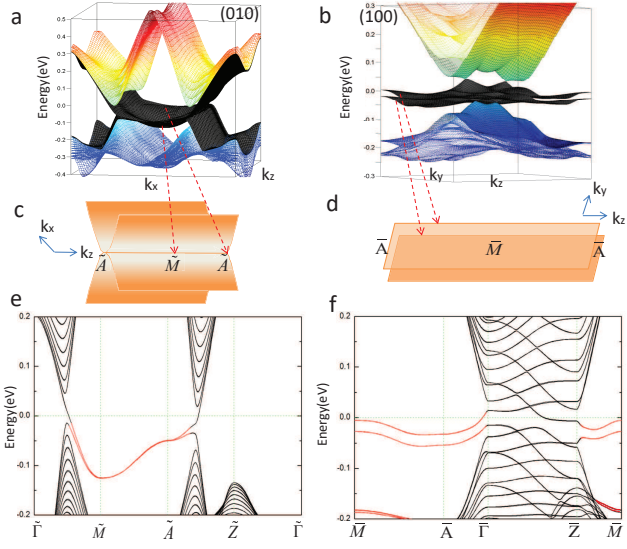}
\par\end{centering}
\caption{(a) The band structures for the PCB-AFM state with the {[}001{]}%
-direction magnetization in the presence of spin-orbit coupling with surface
cuts about the {[}010{]} directions. The thickness of the slab is 20 iron
layers. (b) The band structures for the PCB-AFM state with the {[}001{]}%
-direction magnetization in the presence of spin-orbit coupling with surface
cuts about the {[}100{]} directions. The thickness of the slab is 16 iron
layers. In (a) and (b), the middle heavy-black colored bands are localized
surface bands. (c) and (d) are schematic diagrams of the localized surface
bands which are topologically equivalent to the localized surface bands in
(a) and (b), respectively. (e) and (f) are the surface bands along the
high-symmetry lines in surface BZ shown in Fig. \protect\ref{fig_LDA1}c, and
the localized surface bands are labelled with red color. (e) and (f)
corresponds to (a) and (b) respectively. The thickness of the slab is 60
iron layers in (e) and 40 iron layers in (f).}
\label{fig_band_surface}
\end{figure}

To assess the topological characteristics of the gapped band structure of
tetragonal FeS shown in Fig. \ref{fig_band_LDA}(b), the spectra of surface
states is directly computed. The PCB-AFM order shown in Fig. \ref{fig_LDA1}%
(b) corresponds to a wave vector ($\pi ,\pi /2,0$), thus the surface
commensurate with the PCB-AFM order includes (100) and (010) surfaces. Using
the first-principles calculations, we explicitly demonstrate the presence of
the surface states in a slab geometry in Figs. \ref{fig_band_surface}(a) and %
\ref{fig_band_surface}(b) for (010) and (100) S-terminated surface cuts,
respectively. The results for the Fe-terminated surface cuts are similar. We
extract the localized surface bands and plot the topologically equivalent
schematic diagrams in Figs. \ref{fig_band_surface}(c) and \ref%
{fig_band_surface}(d) to show the key features of the surface states. One
can find that the localized (010) surface bands cross the band gap, which
implies the non-trivial properties as shown in Fig. \ref{fig_band_surface}%
(c), and (e), while the two middle localized surface bands of the (100)
surface open a gap and do not cross the band gap as shown in Fig. \ref%
{fig_band_surface}(d), and (f). The surface spectra definitely demonstrate
the tetragonal FeS with the PCB-AFM order hosts a surface-dependent
nontrivial topological phase.

In the spirit of the principles of symmetry-protected topological phases,
the topological robustness of the surface states is protected by symmetries.
To elucidate the characteristics of the surface states, we first need to
analyze the symmetries owned by (010) and (100) surfaces. Given the PCB-AFM
pattern shown in Fig. \ref{fig_LDA1}(b) with the {[}001{]}-direction
magnetization and taking the mid-point of Fe-Fe bond as origin, the (010)
surface has the glide-plane mirror symmetry $\frac{1}{2}\hat{t}(2\hat{x},0,0)%
\hat{M}_{z}$, and the glide-plane time-reversal symmetry $\frac{1}{2}\hat{t}%
(2\hat{x},0,0)\hat{T}$ with $\hat{T}$ the time-reversal operator; the (100)
surface has only glide-plane time-reversal symmetry $\frac{1}{2}\hat{t}(0,4%
\hat{y},0)\hat{T}$. Note that the fractional translation must be combined
with the point group operators and time-reversal operator to guarantee
system invariant under the combined operations in the presence of
antiferromagnetic order \cite{Mong-PRB-2010,Fang-PRB-2013}. The surface
states can be classified according to the eigen-values of the relevant
symmetry. The representations of symmetry operators upon the surface states
can be constructed as follows: $\frac{1}{2}\hat{t}(2\hat{x},0,0)\hat{M}_{z}%
\mathtt{=}\mathtt{-}ie^{\mathtt{-}ik_{x}}s_{z}$, $\hat{M}_{x}\mathtt{=}%
\mathtt{-}is_{x}$, $\frac{1}{2}\hat{t}(2\hat{x},0,0)\hat{T}\mathtt{=}\mathtt{%
-}ie^{\mathtt{-}ik_{x}}s_{y}\mathcal{K}$, and $\frac{1}{2}\hat{t}(0,4\hat{y}%
,0)\hat{T}\mathtt{=}\mathtt{-}ie^{\mathtt{-}i2k_{y}}s_{y}\mathcal{K}$ with $%
\mathcal{K}$ the complex conjugate operator.

For details, we first discuss the (010) surface. (1) Along the $\tilde{M}%
\mathtt{-}\tilde{A}$ line, the surface bands are a doublet protected by a
product operator $\frac{1}{2}\hat{t}(2\hat{x},0,0)\hat{T}\hat{M}_{z}$, which
results in the pseudo-Kramers degeneracy shown in Fig. \ref{fig_band_surface}%
(e) through the anti-unitary property $[\frac{1}{2}\hat{t}(2\hat{x},0,0)\hat{%
T}\hat{M}_{z}]^{2}\mathtt{=}e^{-i2k_{x}}|_{k_{x}=\pi /2}\mathtt{=}\mathtt{-}%
I $. (2) Along both $\tilde{\Gamma}\mathtt{-}\tilde{M}$ ($k_{z}\mathtt{=}0$)
and $\tilde{Z}\mathtt{-}\tilde{A}$ ($k_{z}\mathtt{=}\pi /c$) lines, $[\frac{1%
}{2}\hat{t}(2\hat{x},0,0)\hat{M}_{z}]^{2}\mathtt{=}$ $-e^{-i2k_{x}}I$ means
two branches for mirror eigenvalues $\pm ie^{-ik_{x}}$. The time-reversal
symmetry enforces the $\frac{1}{2}\hat{t}(2\hat{x},0,0)\hat{M}_{z}$
eigenvalues to be paired as $\{+i,-i\}$ at $k_{x}\mathtt{=}0$ and $\{1,-1\}$
at $k_{x}\mathtt{=}\pi /2$. However, the degenerate points at $\tilde{\Gamma}
$ or $\tilde{Z}$ labeled by $\{+i,-i\}$ merge into the bulk bands and no
edge states survive along the $\tilde{\Gamma}\mathtt{-}\tilde{Z}$ line as
shown in Fig. \ref{fig_band_surface}(e). Fortunately, the degeneracy at the $%
\tilde{M}$ and $\tilde{A}$ points labeled by mirror eigenvalues $\{1,-1\}$
indicates the non-trivial topological properties of the (010) surface states
as shown in Fig. \ref{fig_band_surface}(e). Note that the degeneracy between
the $\tilde{M}$ and $\tilde{A}$ can be slightly gapped by artificially
adding the symmetry-allowed high-order terms without breaking the degeneracy
at the $\tilde{M}$ and $\tilde{A}$ points. As such, the node line connecting
the $\tilde{M}$ and $\tilde{A}$ is reduced into two Dirac nodes at the $%
\tilde{M}$ and $\tilde{A}$ points, and the topological nature of the (010)
surface is characterized by a mirror Chern number $n_{M}\mathtt{=}(n_{1}%
\mathtt{-}n_{-1})/2\mathtt{=}2$ \cite{Teo-PRB-2007,Hsieh-NC-2012}. However,
the symmetry-allowed high-order terms are negligibly small from the
first-principles calculations, because such terms result from the interlayer
couplings along the [001] direction and are beyond the nearest-neighbor
coupling. The constraints from (1) and (2) determine the characteristics of
the (010) surface states. For the (100) surface, along the $\bar{M}\mathtt{-}%
\bar{A}$ line, the product operator has the property of $[\frac{1}{2}\hat{t}%
(0,4\hat{y},0)\hat{T}]^{2}\mathtt{=}\mathtt{-}e^{-i4k_{y}}|_{k_{y}=\pi /4}I%
\mathtt{=}I$. Thus, the in-gap surface states are not degenerate as shown in
Fig. \ref{fig_band_surface}(f). Along the $\bar{\Gamma}\mathtt{-}\bar{Z}$
line, the surface bands merge into the bulk bands as shown in Figs. \ref%
{fig_band_surface}(b) and \ref{fig_band_surface}(f). The remarkable
difference between the (010) and (100) surfaces roots in the strong
nematicity accompanying the PCB-AFM order, which respects the glide-plane
mirror symmetry $\frac{1}{2}\hat{t}(2\hat{x},0,0)\hat{M}_{z}$ of the (010)
surface, but breaks the glide-plane mirror symmetry $\frac{1}{2}\hat{t}(0,4%
\hat{y},0)\hat{M}_{z}$ of the (100) surface. Further combination between the
glide-plane mirror symmetry $\frac{1}{2}\hat{t}(2\hat{x},0,0)\hat{M}_{z}$
and time-reversal symmetry gives the product symmetry $\frac{1}{2}\hat{t}(2%
\hat{x},0,0)\hat{T}\hat{M}_{z}$ which protects the topological crystalline
antiferromagnetic state here.

\section{Discussions}

The non-trivial surface states hosted by tetragonal FeS are determined by
the orientation of magnetization of PCB-AFM order. For a tetragonal crystal,
the spontaneous easy axis or easy plane of the magnetization is determined
by the intrinsic uniaxial anisotropy and tetragonal anisotropy from the
crystallographic structure. Our first-principles calculations show that the
energy difference between the case with the {[}001{]}-direction
magnetization and the case with in-$xy$-plane magnetization is about 0.25
meV/Fe, which is very small. Recent spin-resolved STM measurement has shown
that the magnetization tends to be out-of-$xy$-plane direction near the
surface of bulk Fe$_{1+y}$Te \cite{Torben-NC-2016}. This indicates that some
secondary external effects, such as pressure, mechanical stress, and
alloying, can play an important role in tuning the easy-axis of
magnetization. Indeed, mechanical stress is widely applied in the study of
antiferromagnetic states in iron-based superconductors \cite%
{Chu-Science-2010,Tanatar-PRB-2010}. The strain effect is the
\textquotedblleft inverse\textquotedblright\ of magnetostriction, and the
energy density associated with the strain can be written as $\mathcal{E}%
\mathtt{=}\mathtt{-}\frac{3}{2}\lambda \sigma \cos ^{2}\theta $, where $%
\sigma $ is the stress, $\lambda $ is the magnetostriction constant, and the
angle $\theta $ measures the direction of the magnetization relative to the
direction of the uniform stress. For a positive $\lambda $, the easy-axis is
the {[}001{]}-direction when the stress is along the {[}001{]}-direction.
Thus, modulation of the direction of magnetization supplies a new method to
control the charge and quantum transport of the surface electrons.

One of the remarkable features of tetragonal FeS is the presence of
superconductivity at low temperatures. A consubstantial structure, in which
one side is an superconducting sample and the other side is an topological
crystalline antiferromagnetic state sample, can be fabricated to study the
superconducting proximity effect, leading to topological superconductivity.
In comparison with a heterostructure fabricated by conventional
superconductors and topological insulators or semiconductors\cite%
{Fu-PRL-2008,Lutchyn-PRL-2010}, such a consubstantial structure has many
advantages to eliminate the complexity and unpredictability induced by the
mismatched interface couplings from different materials, and may provide a
platform to explore new physics resulting from the interplay of topology,
magnetism and superconductivity.

In conclusion, an topological crystalline antiferromagnetic state is
proposed to be present accompanying with the PCB-AFM state of tetragonal
FeS, which is protected by the triple fractional translation, mirror
reflection, and time-reversal symmetry. The finding sheds light on exploring
new topological phases protected by non-symmorphic symmetry in
antiferromagnetic materials.

\textbf{Acknowledgements} This work was supported by National Key R\&D Program of China under Grant number: 2017YFA0303201,
NSFC under Grants No. 11674331, No. 11474030, and No. 11625415, 100 Talents Programme of Chinese Academy of Sciences (CAS),
the National Basic Research Program of China under Grant No. 2015CB921103,
the Science Challenge Project under Grant No. JCKY2016212A501, President
Fund of China Academy of Engineering Physics under Grant No. YZJJLX2016010,
and the Research Grant Council of Hong Kong under Grant No. HKU703713P.


\begin{thebibliography}{99}
\bibitem{Hasan-RMP-2010} M. Z. Hasan, and C. L. Kane, Colloquium:
topological insulators. Rev. Mod. Phys. \textbf{82}, 3045-3067 (2010).

\bibitem{Qi-RMP-2011} X.-L. Qi, and S.-C. Zhang, Topological insulators and
superconductors, Rev. Mod. Phys. \textbf{83}, 1057 (2011).

\bibitem{Shen-book-12} S. Q. Shen, Topological insulators: Dirac equation in
condensed matters, (Springer, Berlin, 2012).

\bibitem{Schnyder-PRB-2008} A. P. Schnyder, S. Ryu, A. Furusaki, and A. W.
W. Ludwig, Classification of topological insulators and superconductors in
three spatial dimensions. Phys. Rev. B \textbf{78}, 195125 (2008).

\bibitem{Chen-Science-2012} X. Chen, Z.-C. Gu, Z.-X. Liu, and X.-G. Wen,
Symmetry-protected topological orders in interacting bosonic systems.
Science \textbf{338}, 1604 (2012).

\bibitem{Gu-PRB-2009} Z.-C. Gu, and X.-G. Wen, Tensor-entanglement-filtering
renormalization approach and symmetry-protected topological order. Phys.
Rev. B \textbf{80}, 155131 (2009).

\bibitem{Fu-PRL-2011} L. Fu, Topological crystalline insulators. Phys. Rev.
Lett. \textbf{106}, 106802 (2011).

\bibitem{Hsieh-NC-2012} T. H. Hsieh, H. Lin, J. Liu, W. Duan, A. Bansil, and
L. Fu, Topological crystalline insulators in the SnTe material class. Nat.
Commun. 3, 982 (2012).

\bibitem{Yu-Science-2010} R. Yu, W. Zhang, H. J. Zhang, S. C. Zhang, X. Dai,
and Z. Fang, Quantized anomalous Hall effect in magnetic topological
insulators. Science \textbf{329}, 61 (2010).

\bibitem{Chang-Science-2013} C. Z. Chang, \textit{et al}, Experimental
observation of the quantum anomalous Hall effect in a magnetic topological
insulator, Science \textbf{340}, 167 (2013).

\bibitem{Wang-nature-2016} Z. Wang, A. Alexandradinata, R. J. Cava, and B.
A. Bernevig, Hourglass fermions, Nature \textbf{532}, 189-194 (2016).

\bibitem{Chang-NP-2017} P.-Y. Chang, O. Erten, and P. Coleman, Topological M%
\"{o}bius Kondo insulators, Nat. Phys. in press (2017).

\bibitem{Teo-PRB-2007} J. C. Y. Teo, L. Fu, and C. L. Kane, Surface states
and topological invariants inthree-dimensional topological insulators:
Application to Bi$_{1-x}$Sb$_{x}$. Phys. Rev. B \textbf{78}, 045426 (2008).

\bibitem{Barker-PRL-2016} J. Barker, and O. A. Tretiakov, Static and
Dynamical Properties of Antiferromagnetic Skyrmions in the Presence of
Applied Current and Temperature. Phys. Rev. Lett. \textbf{116}, 147203
(2016).

\bibitem{Zhang-SR-2016} X. Zhang, Y. Zhou, and M. Ezawa, Antiferromagnetic
Skyrmion: Stability, Creation and Manipulation. Sci. Rep. \textbf{6}, 24795
(2016).

\bibitem{Jungwirth-NN-2016} T. Jungwirth, X. Marti, P. Wadley, and J.
Wunderlich, Antiferromagnetic spintronics. Nat. Nano. \textbf{11}, 231
(2016).

\bibitem{Lai-JACS-2015} X. Lai, H. Zhang, Y. Wang, X. Wang, X. Zhang, J.
Lin, and F. Huang, Observation of Superconductivity in Tetragonal FeS, J.
Am. Chem. Soc. \textbf{137}, 10148 (2015).

\bibitem{Borg-PRB-2016} C. K. H. Borg, X. Zhou, C. Eckberg, D. J. Campbell,
S. R. Saha, J. Paglione, and E. E. Rodriguez, Strong anisotropy in nearly
ideal tetrahedral superconducting FeS single crystals, Phys. Rev. B \textbf{%
93}, 094522 (2016).

\bibitem{Kuhn-Ph-2016} S.J. Kuhn, M.K. Kidder, W.M. Chance, C. dela Cruz,
M.A. McGuire, D.S. Parker, L. Li, L. Debeer-Schmitt, J. Ermentrout, K.
Littrell, M.R. Eskildsen, A.S. Sefat, FeS: Structure and Composition
Relations to Superconductivity and Magnetism, Physica C: Superconductivity
and its applications \textbf{534}, 29-36 (2017).

\bibitem{Holenstein-PRB-2016} S. Holenstein, U. Pachmayr, Z. Guguchia, S.
Kamusella, R. Khasanov, A. Amato, C. Baines, H.-H. Klauss, E. Morenzoni, D.
Johrendt, and H. Luetkens, Coexistence of low-moment magnetism and
superconductivity in tetragonal FeS and suppression of $T_{c}$ under
pressure Phys. Rev. B \textbf{93}, 140506 (2016).

\bibitem{Kirschner-PRB-2016} F. K. K. Kirschner, F. Lang, C. V. Topping, P.
J. Baker, F. L. Pratt, S. E. Wright, D. N. Woodruff, S. J. Clarke, and S. J.
Blundell, Robustness of superconductivity to competing magnetic phases in
tetragonal FeS, Phys. Rev. B \textbf{94}, 134509 (2016).

\bibitem{vasp} G. Kresse, and J. Furthm\"{u}ller, Efficient iterative
schemes for \textit{ab initio} total-energy calculations using a plane-wave
basis set. Phys. Rev. B \textbf{54}, 11169 (1996).

\bibitem{paw1} P. E. Bl\"{o}chl, Projector augmented-wave method. Phys. Rev.
B \textbf{50}, 17953 (1994).

\bibitem{paw2} G. Kresse, and D. Joubert, From ultrasoft pseudopotentials to
the projector augmented-wave method. Phys. Rev. B \textbf{59}, 1758 (1999).

\bibitem{pbe1} J.~P. Perdew, K. Burke, and M. Ernzerhof, Generalized
Gradient Approximation Made Simple. Phys. Rev. Lett. \textbf{77}, 3865
(1996).

\bibitem{vdw} J. Klime\v{s}, D. R. Bowler, and A. Michaelides, Van der Waals
density functionals applied to solids. Phys. Rev. B \textbf{83}, 195131
(2011).

\bibitem{Man-njp-2017} Haoran Man ,\textit{et. al.}, Spin excitations and
the Fermi surface of superconducting FeS, njp Quantum Materials 2, 14 (2017).

\bibitem{Sun-NC-2016} J. P. Sun \textit{et. al.}, Dome-shaped magnetic order
competing with high-temperature superconductivity at high pressures in FeSe,
Nat. Commn. 7, 12146 (2016).

\bibitem{Sun-PRL-2017} J. P. Sun \textit{et. al.}, High-$T_{c}$
Superconductivity in FeSe at High Pressure: Dominant Hole Carriers and
Enhanced Spin Fluctuations, Phys. Rev. Lett. \textbf{118}, 147004 (2017).
\textit{\ }

\bibitem{Cao-PRB-2015} H.-Y. Cao, S. Chen, H. Xiang, and X.-G. Gong,
Antiferromagnetic ground state with pair-checkerboard order in FeSe, Phys.
Rev. B \textbf{91}, 020504(R) (2015).

\bibitem{Liu-PRB-2016} K. Liu, Z.-Y. Lu, and T. Xiang, Nematic
antiferromagnetic states in bulk FeSe, Phys. Rev. B \textbf{93}, 205154
(2016).

\bibitem{Glasbrenner-NP-2015} J. K. Glasbrenner, I. I. Mazin, H. O. Jeschke,
P. J. Hirschfeld, R. M. Fernandes, and R. Valent\'{\i}, Effect of magnetic
frustration on nematicity and superconductivity in iron chalcogenides, Nat.
Phys. \textbf{11}, 953-958 (2015).

\bibitem{Hu-PRB-2012} J. Hu, B. Xu, W. Liu, N.-N. Hao, and Y. Wang, Unified
minimum effective model of magnetic properties of iron-based
superconductors, Phys. Rev. B \textbf{85}, 144403 (2012).

\bibitem{Richard-PRL-2010} P. Richard, \textit{et al. }Observation of Dirac
Cone Electronic Dispersion in BaFe$_{2}$As$_{2}$, Phys. Rev. Lett. \textbf{%
104}, 137001 (2010).

\bibitem{Hao-PRX-2014} N. Hao, and J. Hu, Topological Phases in the
Single-Layer FeSe, Phys. Rev. X \textbf{4}, 031053 (2014).

\bibitem{Hao-PRB-2015} N. Hao,and S.-Q. Shen, Topological superconducting
states in monolayer FeSe/SrTiO$_{3}$, Phys. Rev. B \textbf{92}, 165104
(2015).

\bibitem{Mong-PRB-2010} R. S. K. Mong, A. M. Essin, and J. E. Moore, Phys.
Rev. B \textbf{81}, 245209 (2010).

\bibitem{Fang-PRB-2013} C. Fang, M. J. Gilbert, and B. A. Bernevig, Phys.
Rev. B \textbf{88}, 085406 (2013).

\bibitem{Torben-NC-2016} T. H\"{a}nke, \textit{et al.} Reorientation of the
diagonal double-stripe spin structure at Fe$_{1+y}$Te bulk and thin-film
surfaces, Nat. Comms. \textbf{8}, 13939 (2016).

\bibitem{Chu-Science-2010} J.-H. Chu, \textit{et al.} In-plane resistivity
anisotropy in an underdoped iron arsenide superconductor, Science \textbf{329%
}, 824 (2010).

\bibitem{Tanatar-PRB-2010} M. A. Tanatar, \textit{et al.} Uniaxial-strain
mechanical detwinning of CaFe$_{2}$As$_{2}$ and BaFe$_{2}$As$_{2}$ crystals:
Optical and transport study, Phys. Rev. B \textbf{81}, 184508 (2010).

\bibitem{Fu-PRL-2008} L. Fu, and C. L. Kane, Superconducting Proximity
Effect and Majorana Fermions at the Surface of a Topological Insulator,
Phys. Rev. Lett. \textbf{100}, 096407 (2008).

\bibitem{Lutchyn-PRL-2010} R. M. Lutchyn, J. D. Sau, and S. D. Sarma,
Majorana Fermions and a Topological Phase Transition in
Semiconductor-Superconductor Heterostructures, Phys. Rev. Lett. \textbf{105}%
, 077001 (2010).
\end{thebibliography}
\end{document}